%% file: Conference-LaTeX-template_10-17-19/issre2025.tex
\documentclass[conference]{IEEEtran}
\IEEEoverridecommandlockouts
\usepackage{cite}
\usepackage{amsmath,amssymb,amsfonts}
\usepackage{algorithmic}
\usepackage{graphicx}
\usepackage{textcomp}
\usepackage{xcolor}
\usepackage{cite}
\usepackage{amsmath,amssymb,amsfonts}
\usepackage{algorithmic}
\usepackage{graphicx}
\usepackage{textcomp}
\usepackage{xcolor}
\usepackage{times}
\usepackage{soul}
\usepackage{url}
\usepackage[hidelinks]{hyperref}
\usepackage[utf8]{inputenc}
\usepackage[small]{caption}
\usepackage{graphicx}
\usepackage{amsmath}
\usepackage{amsthm}
\usepackage{booktabs}
\usepackage{algorithm}
\usepackage{algorithmic}
\usepackage[switch]{lineno}
\usepackage{multirow}
\usepackage{subfigure}
\usepackage{booktabs}
\usepackage{multirow}
\usepackage{graphicx}
\usepackage{ulem} 
\newcommand{\name}{\textit{TShape}~}

\def\BibTeX{{\rm B\kern-.05em{\sc i\kern-.025em b}\kern-.08em
    T\kern-.1667em\lower.7ex\hbox{E}\kern-.125emX}}

\usepackage{changes}

\begin{document}



\title{TShape: Rescuing Machine Learning Models from Complex Shapelet Anomalies}

\author{
    \IEEEauthorblockN{
        Hang Cui\IEEEauthorrefmark{1}\IEEEauthorrefmark{2}, 
        Jingjing Li\IEEEauthorrefmark{1*}, 
        Haotian Si\IEEEauthorrefmark{1}\IEEEauthorrefmark{2}, 
        Quan Zhou\IEEEauthorrefmark{1}\IEEEauthorrefmark{2}, 
        Changhua Pei\IEEEauthorrefmark{1*}\IEEEauthorrefmark{3},
        Gaogang Xie\IEEEauthorrefmark{1}, 
        Dan Pei\IEEEauthorrefmark{4}, 
    }
    \IEEEauthorblockA{\IEEEauthorrefmark{1}Computer Network Information Center, Chinese Academy of Sciences, \{cuihang, ljj, htsi, zhouquan, chpei, xie\}@cnic.cn}
    \IEEEauthorrefmark{2}University of the Chinese Academy of Sciences, \\
    \IEEEauthorrefmark{3}Hangzhou Institute for Advanced Study, University of Chinese Academy of Sciences \\
    \IEEEauthorblockA{\IEEEauthorrefmark{4}Tsinghua University, peidan@tsinghua.edu.cn}
}


\maketitle

\begin{abstract}
Time series anomaly detection (TSAD) is critical for maintaining the reliability of modern IT infrastructures, where complex anomalies frequently arise in highly dynamic environments. In this paper, we present \textit{TShape}, a novel framework designed to address the challenges in industrial time series anomaly detection. Existing methods often struggle to detect shapelet anomalies that manifest as complex shape deviations, which appear obvious to human experts but prove challenging for machine learning algorithms. \name introduces a patch-wise dual attention mechanism with multi-scale convolution to model intricate sub-sequence variations by balancing local, fine-grained shape features with global contextual dependencies. Our extensive evaluation on five diverse benchmarks demonstrates that \name outperforms existing state-of-the-art models, achieving an average 10\% F1 score improvement in anomaly detection. Additionally, ablation studies and attention visualizations confirm the essential contributions of each component, highlighting the robustness and adaptability of \name to complex shapelet shapes in time series data.
\end{abstract}

\begin{IEEEkeywords}
Time Series Shapelet, Anomaly Detection, Patch-wise Dual Attention
\end{IEEEkeywords}

\input{Conference-LaTeX-template_10-17-19/chapters/01-content/10-introduction}
\input{Conference-LaTeX-template_10-17-19/chapters/01-content/20-background}
\input{Conference-LaTeX-template_10-17-19/chapters/01-content/30-approach}

\input{Conference-LaTeX-template_10-17-19/chapters/01-content/40-EVALUATION}

\input{Conference-LaTeX-template_10-17-19/chapters/01-content/50-CONCLUSION}
\input{Conference-LaTeX-template_10-17-19/chapters/01-content/Acknowledgement}

\bibliographystyle{IEEEtran}
\bibliography{issre2025}
\end{document}

%% file: Conference-LaTeX-template_10-17-19/chapters/01-content/10-introduction.tex
\section{Introduction}\label{sec:intro}

Time Series Anomaly Detection (TSAD) is critical for ensuring the operational health of IT infrastructure and the reliability of software systems.  IT operations engineers must continuously monitor key time-series metrics, such as response times or success rates, derived from the vast volumes of data generated daily by IT infrastructure~\cite{liu2025opseval}.  This monitoring is essential to maintain service quality and user satisfaction by promptly identifying system failures.  The field has evolved significantly, transitioning from manual monitoring and rule-based statistical methods to modern deep learning approaches, particularly those based on prediction-observation comparisons.  Recent large-scale service outages at major cloud providers (e.g. Microsoft ~\cite{Microsoft}, Google~\cite{Google}, and Alibaba Cloud~\cite{Alibaba}) underscore the urgent need for more effective TSAD solutions.
\begin{figure}[tbp]
  \centering
  \includegraphics[width=0.47\textwidth]{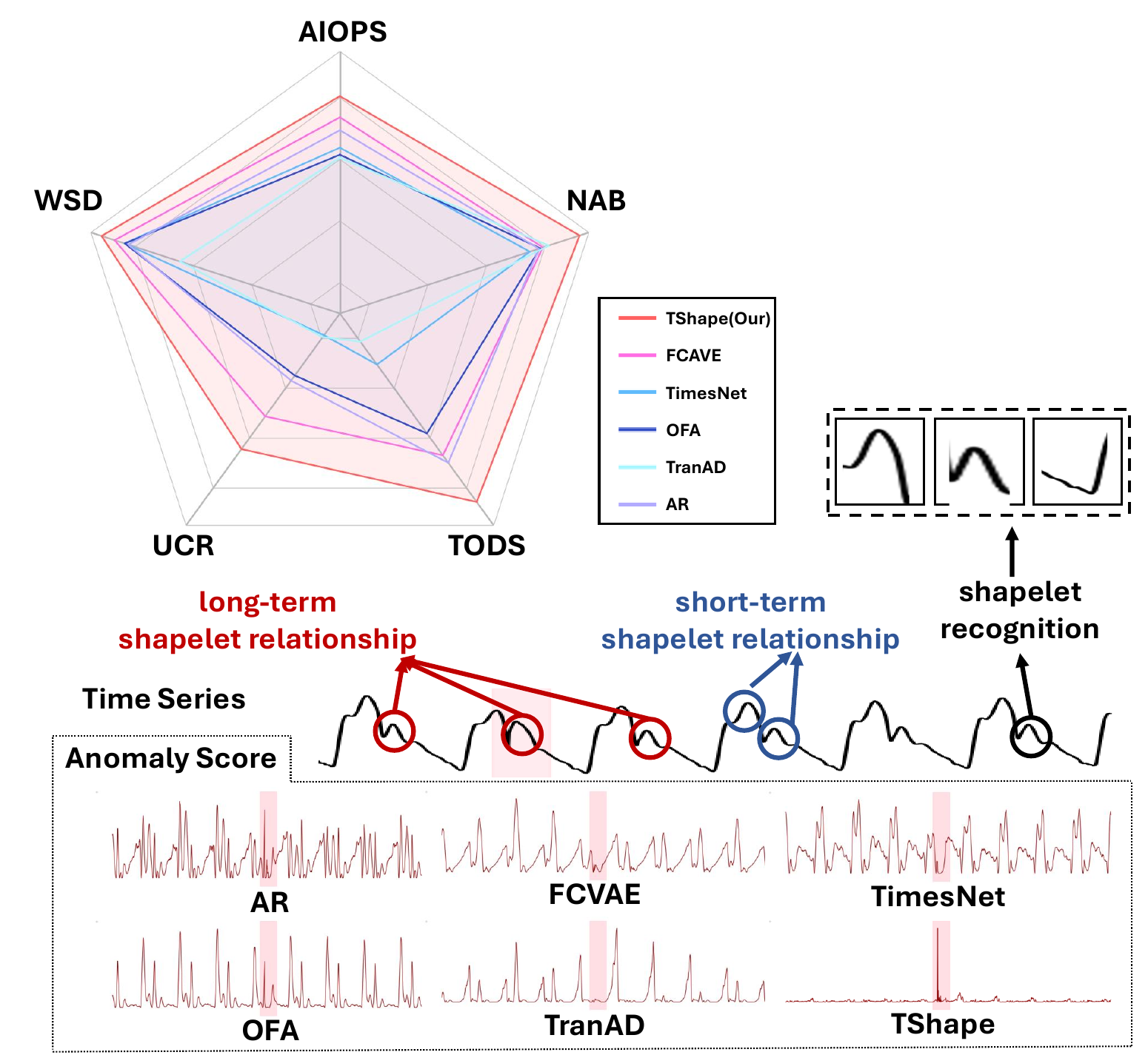}
    \caption{Performance evaluation of \name. (a) Top section: Detection effectiveness of time series anomaly detection methods across datasets (see Table~\ref{table:results} for quantitative metrics). (b) Bottom section: Illustration of anomaly scoring where the black curve represents a sample time series, the pink background denotes ground-truth anomaly intervals, and red curves indicate time-point anomaly scores generated by each method.}  \label{figure:case}
\end{figure}

Recent years have witnessed significant advances in deep learning-based time series anomaly detection, with numerous novel methods emerging~\cite{si2024timeseriesbench}. For instance, AnomalyTransformer~\cite{xu2021AnomalyTransformer} leverages the Transformer architecture to identify anomalies by quantifying attention distribution divergence between normal and anomalous data. FCVAE~\cite{wang2024revisiting} decomposes time series into multi-scale frequency components, achieving state-of-the-art performance through joint modeling of temporal and spectral patterns. 


Although there are many methods available, existing time series anomaly detection methods primarily focus on modeling the relationships between 'points', either long-term or short-term. They fail to sufficiently analyze the interrelationships between shapelets. This inadequacy misaligns with how human experts assess anomalies, leading to situations where machine learning algorithms cannot detect anomalies that are easily identifiable by human experts. An example is illustrated in Fig.\ref{figure:case}, where the middle section shows a segment of the time series. The anomalous segment, highlighted with a pink background, is relatively easy for human experts to identify. This is because: In each cycle, there are two small peaks (shapelet recognition), where the second, smaller peak (highlighted in red) exhibits a different convex shape at the same position during the anomalous period compared to other cycles (long-term shapelet relationship). Meanwhile, the relative magnitude between the two peaks during the anomalous period is different; the amplitude difference between the two peaks becomes smaller during the anomalous period compared to other periods where the difference is larger (short-term shapelet relationship). It can be seen that human anomaly detection is at the shapelet level, whereas existing machine learning methods attempt to detect at the point level. Shapelet recognition is a trending topic in the image domain, such as ShapeFormer~\cite{yan2022shapeformer}, but there isn't currently an effective method for modeling time series shapelets; existing methods mainly compare the distance between two segments, like DAMP~\cite{lu2022DAMP} and PatchAD~\cite{zhong2025patchad}. Making machine learning algorithms think like operational experts poses two challenges: 

\begin{itemize}


\item \textbf{How to identify anomalous shapelets among diverse shapelets:}  Real-world anomalies exhibit highly diverse shapes. Developing a unified model capable of learning robust representations to recognize anomalies across a wide spectrum of morphological patterns, especially such as the complex shape deviations shown in Fig.~\ref{figure:case}, remains a critical and open challenge in industrial TSAD.

\item \textbf{How to simultaneously consider short-term and long-term shapelet relationships:} Local shapelet relationships represent detailed patterns, while long-distance shapelet relationships are used to check for period violations, which are more robust than point-based periodicity checks (e.g., FCVAE~\cite{wang2024revisiting}). Effectively modeling this relationship and integrating fine-grained local and global relationships is a new problem.

\end{itemize}

To tackle these challenges in industrial time series anomaly detection, we present \textit{TShape}: 
To address Challenge 1, \name implements patch-based modeling, dividing the input sequence into segments to capture intricate sub-sequence variations that are often neglected by point-based models. Besides, we utilize multi-scale convolution to extract rich local descriptors that accurately represent complex temporal features through the application of parallel multi-scale convolutions.
To address Challenge 2, we introduce a patching dual‑attention mechanism, in which local attention models intra-patch dependencies within each patch, global attention captures inter‑patch relationships, and a learnable gating unit adaptively fuses these signals to balance local complex shapes and global relationships. 
Our key contributions to the field of Time Series Anomaly Detection are as follows:

\begin{itemize}
    \item We introduce Multi-scale Convolution to model complex sub-sequence variations, addressing the critical limitation of inadequate utilization of localized shape information in existing TSAD methods.

    \item We develop the Patch-wise Dual‑Attention mechanism with a gated fusion of local and global attention streams, ensuring that both intra‑patch and inter‑patch dependencies are effectively leveraged.

    \item We conduct comprehensive evaluations on five widely-used benchmark datasets, demonstrating that \name surpasses current SOTA methods in the TSAD benchmarks. Our code is publicly available at  https://github.com/CSTCloudOps/TShape.
\end{itemize}

%% file: Conference-LaTeX-template_10-17-19/chapters/01-content/20-background.tex
\section{BACKGROUND AND RELATED WORKS}\label{sec:background}

\subsection{Problem Statement}

We consider the task of univariate time series anomaly detection. Let $X = \{x_t\}_{t=1}^T, \ x_t \in \mathbb{R}$ be an observed time series of length \(T\), where each \(x_t\) (e.g. CPU utilization or response time) reflects the system’s state at time \(t\).  The underlying system generating $X$ is assumed to operate predominantly under a normal regime, characterized by inherent temporal patterns reflecting healthy system behavior. However, at certain unknown time points, the system may experience anomalous events (e.g. hardware failures, software bugs, network attacks), causing subsequences within $X$ to deviate significantly from this normal behavior. The goal of TSAD is to learn a model $F$ from only normal training data that assigns an anomaly score $s_t\in R$ to each time step $t$ in a new, potentially anomalous sequence $X=\{x_{t-T},x_{t-T+1},...,x_{t}\}$. A higher $s_t$ indicates a higher probability that the observation $x_t$ is anomalous.

\subsection{Related Work}

Time series anomaly detection methods can be broadly categorized into three paradigms~\cite{si2024timeseriesbench}: Statistical Methods, Prediction-based Methods and Reconstruction-based Methods.

\textit{Statistical Methods}: Sub-LOF~\cite{breunig2000lof} identifies local outliers by comparing the density deviation of a point relative to its neighbors.  SAND~\cite{boniol2021sand} employs temporal shape-based clustering to isolate anomalous segments.  MatrixProfile~\cite{zhu2018matrix} computes the minimum Euclidean distance between subsequences and their nearest neighbors in a sliding window.

\textit{Prediction-based Methods}: These methods forecast future values and trigger alarms upon significant prediction errors.  Representative works include AR~\cite{rousseeuw2003AR} and LSTMAD~\cite{malhotra2015LSTMAD}, which uses LSTM networks to capture temporal dependencies. TimesNet~\cite{wu2022timesnet} converts time series into 2D space via frequency folding and processes them with vision-inspired backbones.    OFA~\cite{zhou2023OFA} adapts LLM for time series but inherits its global bias, blurring anomaly-localizing details.

\textit{Reconstruction-based Methods}: These methods learn normal patterns through encoding-decoding, and large reconstruction errors indicate anomalies. Basic Autoencoders(AE~\cite{ng2011AE} and EncDecAD~\cite{malhotra2016EncDecAD}) learn compressed representations of normal data.   TranAD\cite{tuli2022tranad} combines transformers with adversarial training for robust reconstruction. In pursuit of more expressive probabilistic models, recent methods leverage variational inference to capture uncertainty in latent representations:  FCVAE~\cite{wang2024revisiting} advances this line of work by decomposing time series into frequency components and modeling each via a variational framework. 

%% file: Conference-LaTeX-template_10-17-19/chapters/01-content/30-approach.tex
\section{APPROACH}\label{sec:APPROACH}

\begin{figure*}[tbp]
  \centering
  \includegraphics[width=0.97\textwidth]{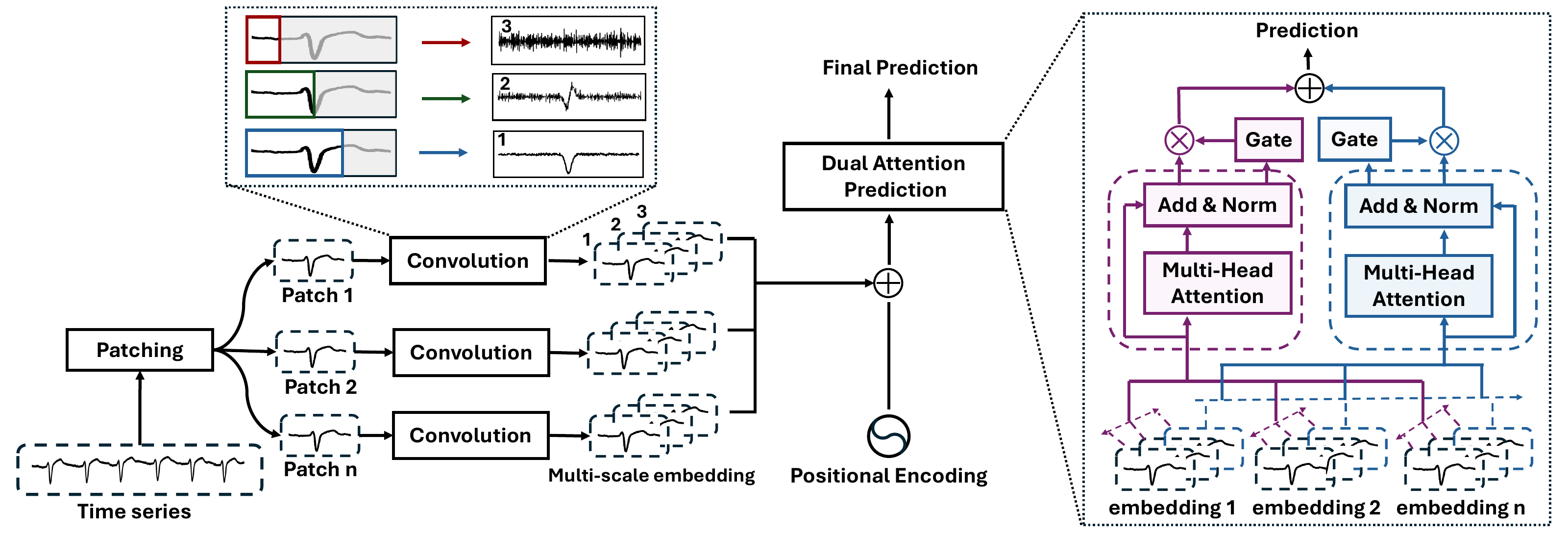}
  \caption{Overview of \name.}
  \label{figure:framework}

\end{figure*}

\subsection{Overview}

As shown in Fig.\ref{figure:framework}, the architecture of \name consists of three main components: 1)Multi-scale Convolution:  Given a univariate time series, \name first splits the input window into patches and applies multi‑scale convolution within each patch to extract rich local features. 2)Patch-wise Positional Encoding: Ensuring the model is aware of each patch temporal position, \name injects explicit patch‑order information via learnable embeddings.  3)Patch-wise Dual‑Attention: In order to capture both local‑grained and long‑range dependencies, \name focus on both intra‑patch and inter‑patch attention, dynamically fusing by gating network.

\subsection{Multi-scale Convolution}

To capture the temporal shapes of individual patches, \name extracts multi-scale local features from each patch by Multi-scale Convolution. Time series data often exhibits patterns at different temporal resolutions, which cannot be captured by a single convolution kernel. By employing multiple convolution kernels of different sizes, \name is able to extract features at varying time scales, which is essential for handling series with multi-scale behavior.
Each patch \(\mathbf{p}_i \in \mathbb{R}^s\) is processed using parallel 1D convolutions with different kernel sizes. \name lets the set of kernel sizes be \( \mathcal{K} = \{k_1, k_2, \dots, k_M\} \), where each kernel $k_i$ captures different aspects of the patch’s temporal behavior.

\[
    \mathbf{h}_i^{(k)} = \mathrm{Conv1D}_k(\mathbf{p}_i) \in \mathbb{R}^{C_m \times s}
\]
    
The output of each convolution is pooled using global average pooling to obtain a feature vector that summarizes the patch's behavior at that scale:
\[
    \mathbf{z}_i^{(k)} = \mathrm{GAP}(\mathbf{h}_i^{(k)}) \in \mathbb{R}^{C_m}
\]
    
After pooling, the multi-scale features are concatenated across all kernel sizes:
    \[
    \mathbf{z}_i = [\mathbf{z}_i^{(k_1)}, \mathbf{z}_i^{(k_2)}, \dots, \mathbf{z}_i^{(k_M)}] \in \mathbb{R}^{C}
    \]
    
This aggregation allows the model to capture dependencies of different scales within each patch. The fused features are then passed through a batch normalization layer and a GELU activation function:
    \[
    \mathbf{u}_i = \mathrm{GELU}(\mathrm{BN}([\mathbf{z}_1, \dots, \mathbf{z}_P]^\top)) \in \mathbb{R}^{P \times C}
    \]
    
The output \(\mathbf{u}_i\) represents the processed features for each patch for further sequence modeling.

\subsection{Patch-wise Positional Encoding}

To ensure awareness of each patch’s temporal position, \name injects explicit patch‑order information via a learnable embedding. Without Patch-wise Positional Encoding, the model would treat each patch as independent of its position within the sequence. Given the convolutional feature matrix
\(\mathbf{U}\in\mathbb{R}^{P\times C}\), we introduce a positional embedding
\(\mathbf{E}\in\mathbb{R}^{P\times C}\) parameterized as
\(\mathbf{E} = \{\mathbf{e}_1, \dots, \mathbf{e}_P\}\). The encoded features are
\[
    \mathbf{V} = \mathbf{U} + \mathbf{E}, 
    \quad \mathbf{E}\sim\mathcal{N}(0,1).
\]
This addition allows the model to distinguish patches based on their order, which is critical for preserving temporal structure in subsequent attention layers.

\subsection{Patch-wise Dual‑Attention}

To capture both local feature shapes and long range dependencies, \name applies dual self‑attention including local attention and global attention over the sequence of patches \(\mathbf{V} \in \mathbb{R}^{P \times C}\). 
The local attention mechanism focuses on capturing intra-patch dependencies by applying multi-head attention across the intra-patch features of the patches. This allows the model to emphasize important features within each patch. For patch features \(\mathbf{V}\), \name first reshape \(\mathbf{V}\) to \(\mathbb{R}^{P\times C}\!\to\!\mathbb{R}^{C\times P}\) and apply multi‑head attention:
    \[
        \tilde{\mathbf{L}} = \mathrm{MHA}_{\mathrm{local}}(\mathbf{V}^\top,\mathbf{V}^\top,\mathbf{V}^\top)^\top + \mathbf{V}.
    \]

This operation emphasizes important features within each patch. The second attention mechanism focuses on capturing the relationships between patches by applying MHA across the patches. \name capture interactions across patches modeling by
    \[
        \tilde{\mathbf{G}} = \mathrm{MHA}_{\mathrm{global}}(\mathbf{V},\mathbf{V},\mathbf{V}) + \mathbf{V},
    \]
This allows the model to learn how patches interact over time, capturing long-range dependencies. The outputs of the local and global attention mechanisms are fused using a learnable gate:
    \[
    g = \sigma\left([\tilde{\mathbf{L}}; \tilde{\mathbf{G}}]W_g + b_g\right)
    \]

The fused features are then combined and outputted:
    \[
    \mathbf{H} = g \odot \tilde{\mathbf{L}} + (1 - g) \odot \tilde{\mathbf{G}}
    \]

%% file: Conference-LaTeX-template_10-17-19/chapters/01-content/40-EVALUATION.tex
\section{EVALUATION}\label{sec:EVALUATION}

\begin{table*}[ht]
\centering
\caption{Performance comparison across five datasets.}
\resizebox{\textwidth}{!}{
\begin{tabular}{lcccccccccccccc}
\toprule
\multirow{2}{*}{\textbf{Method}} & \multicolumn{2}{c}{\textbf{AIOPS}} & \multicolumn{2}{c}{\textbf{NAB}} & \multicolumn{2}{c}{\textbf{TODS}} & \multicolumn{2}{c}{\textbf{UCR}} & \multicolumn{2}{c}{\textbf{WSD}} & \multirow{2}{*}{\textbf{Avg F1-E}} \\
\cmidrule(lr){2-3} \cmidrule(lr){4-5} \cmidrule(lr){6-7} \cmidrule(lr){8-9} \cmidrule(lr){10-11}
 & F1 & F1-E & F1 & F1-E & F1 & F1-E & F1 & F1-E & F1 & F1-E & \\
\midrule
SubLOF & 0.7273 & 0.2805 & 0.9787 & 0.6221 & 0.7997 & 0.4795 & \textbf{0.8811} & 0.4917 & 0.8683 & 0.6585 & 0.5065 \\
SAND & 0.2710 & 0.0397 & 0.7007 & 0.1179 & 0.5372 & 0.1879 & 0.7489 & \uline{0.5250} & 0.1761 & 0.0839 & 0.1909 \\
MatrixProfile & 0.1915 & 0.0171 & 0.7873 & 0.0567 & 0.5284 & 0.1389 & 0.7992 & 0.4438 & 0.1233 & 0.0279 & 0.1369 \\
AR & 0.8775 & 0.6944 & \textbf{0.9982} & 0.7800 & 0.8312 & \uline{0.6987} & 0.7897 & 0.3693 & 0.9624 & 0.8225 & 0.6730 \\
LSTMAD & \textbf{0.9296} & \uline{0.7671} & 0.9912 & 0.7852 & 0.8226 & 0.6582 & 0.7894 & 0.4043 & \textbf{0.9871} & \uline{0.9009} & 0.7031 \\
AE & 0.8778 & 0.6743 & 0.9921 & 0.7738 & 0.8402 & 0.6404 & 0.7074 & 0.3606 & 0.9809 & 0.8851 & 0.6668 \\
EncDecAD & 0.9015 & 0.7022 & 0.9900 & 0.7700 & 0.7346 & 0.6156 & 0.6685 & 0.2541 & 0.9826 & 0.8945 & 0.6473 \\
SRCNN & 0.4069 & 0.1338 & 0.9079 & 0.4275 & 0.6298 & 0.2066 & 0.7496 & 0.2209 & 0.4010 & 0.1221 & 0.2222 \\
AT & 0.6235 & 0.3124 & 0.9757 & 0.6117 & 0.5249 & 0.2123 & 0.7251 & 0.2076 & 0.4414 & 0.2462 & 0.3180 \\
TFAD & 0.6020 & 0.2638 & 0.8609 & 0.4979 & 0.6686 & 0.3751 & 0.6456 & 0.2846 & 0.8572 & 0.7113 & 0.4265 \\
TranAD & 0.7696 & 0.6051 & 0.9975 & 0.8133 & 0.5131 & 0.2127 & 0.6101 & 0.1946 & 0.7616 & 0.6468 & 0.4945 \\
Donut & 0.8595 & 0.6299 & 0.9834 & 0.6938 & 0.8655 & 0.6826 & 0.7514 & 0.3580 & 0.9637 & 0.8326 & 0.6394 \\
FCVAE & 0.9183 & 0.7364 & 0.9948 & \uline{0.7933} & \uline{0.8658} & 0.6689 & 0.8357 & 0.5126 & 0.9689 & 0.8695 & \uline{0.7161} \\
TimesNet & 0.8018 & 0.6379 & 0.9875 & 0.7488 & 0.6234 & 0.3048 & 0.6210 & 0.1873 & 0.9354 & 0.8358 & 0.5429 \\
OFA & 0.8810 & 0.6150 & 0.9962 & 0.7841 & 0.6928 & 0.5811 & 0.7286 & 0.3489 & 0.9564 & 0.8344 & 0.6327 \\
FITS & 0.9108 & 0.6324 & 0.9962 & 0.7447 & 0.7543 & 0.4873 & 0.7567 & 0.3284 & 0.9705 & 0.8377 & 0.6061 \\
\textbf{\name} & \uline{0.9263} & \textbf{0.8049} & \textbf{0.9982} & \textbf{0.9186} & \textbf{0.9085} & \textbf{0.8561} & \uline{0.8493} & \textbf{0.5915} & \uline{0.9829} & \textbf{0.9137} & \textbf{0.8170} \\
\bottomrule
\end{tabular}
}
\label{table:results}

\end{table*}



\subsection{Experimental Settings}

\textbf{Datasets: }To ensure comprehensive coverage of anomaly distributions, we have integrated five meticulously annotated datasets spanning diverse application domains:

\begin{itemize}
  \item \textit{AIOPS~\cite{AIOPS}:} Sourced from five leading Internet firms (Sogou, eBay, Baidu, Tencent, Alibaba), this multidimensional collection comprises system logs, resource metrics, and event traces. It challenges models with evolving distributions, and heterogeneous anomalies ranging from hardware faults to security breaches.
  \item \textit{WSD~\cite{WSD}:} Recorded at sub‑second intervals from production web services at Baidu, Sogou, and eBay, WSD offers 210 annotated series of KPIs such as latency and error rates. Its high sampling rate and bursty volatility test a detector’s responsiveness under rapid fluctuations.
  \item \textit{UCR~\cite{UCR}:} A canonical repository of 203 time-series across domains (such as power‑grid, medical sensors, industrial IoT), each containing a single expert‑verified anomaly interval. UCR measures a model’s generalization across distinct domains and anomaly types.
  \item \textit{TODS~\cite{TODS}:} A synthetic suite in which anomalies are injected with precise control over seasonality, trend, and noise parameters. Its ground‑truth clarity and tunable complexity enable incisive analysis of design components.
  \item \textit{NAB~\cite{NAB}:} Streaming data from real-world AWS cloud metrics, social media activity, and IoT sensors, augmented with synthetic sequences. NAB reflects operational detection scenarios where real‑time processing and hybrid anomaly sources coexist.
\end{itemize}

Each time series in these datasets is treated independently: we train a separate \name instance per univariate sequence and evaluate on its held‑out test split. To evaluate anomaly detection capabilities of \name and ensure a comprehensive evaluation, our training and testing protocol obey the time series benchmark EASYTSAD\footnote{\url{https://adeval.cstcloud.cn/}}.

\textbf{Baselines: }We compare \name against sixteen state‑of‑the‑art methods: SubLOF~\cite{breunig2000lof}, SAND~\cite{boniol2021sand}, MatrixProfile~\cite{zhu2018matrix}, AR~\cite{rousseeuw2003AR}, LSTMAD~\cite{malhotra2015LSTMAD}, AE~\cite{ng2011AE}, EncDecAD~\cite{malhotra2016EncDecAD}, SRCNN~\cite{ren2019SRCNN}, AnomalyTransformer~\cite{xu2021AnomalyTransformer}, TFAD~\cite{zhang2022tfad}, TranAD~\cite{tuli2022tranad}, Donut~\cite{Metric_Best_f1}, FCVAE~\cite{wang2024revisiting}, TimesNet~\cite{wu2022timesnet}, OFA~\cite{zhou2023OFA}and FITS~\cite{xu2023fits}. For each baseline, we use recommended hyperparameters from the original papers.

\begin{figure}[htbp]
  \centering
  \includegraphics[width=0.47\textwidth]{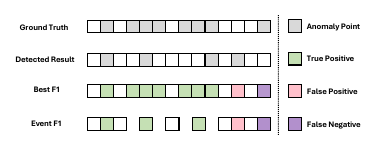}
  \caption{Introduction of metrics.}
  \label{figure:metrics}

\end{figure}

\textbf{Metrics: }To mitigate the inherent threshold selection bias in anomaly detection systems~\cite{Metric_Best_f1}, we employ the Best F1 score as our primary optimization metric. However, prior research~\cite{Best_f1_inflation,Metric_Best_f1} indicates that this conventional metric is susceptible to artificial score inflation.   This inflation stems from the redundant point-wise counting of consecutive anomalies occurring within extended anomalous events.   Recognizing that practical anomaly detection scenarios necessitate identifying coherent anomalous events rather than isolated outlier points, we utilize the Event F1 score~\cite{si2024timeseriesbench}.  As shown in Fig.\ref{figure:metrics}, this metric evaluates segment-level detection accuracy by treating continuous anomalous intervals as single events, effectively decoupling the influence of event duration from the assessment of detection capability.

\subsection{Overall Performance}

Table~\ref{table:results} presents a rigorous comparison of \name against 16 state-of-the-art baselines across five diverse anomaly detection datasets. \name achieves the highest average F1 score of 0.9330 and Event F1 score of 0.8170, significantly outperforming all competitors – a 10\% absolute improvement over the nearest baseline (FCVAE, 0.7161). This demonstrates accuracy of \name  in detecting anomalous events. 

Cross-Dataset Robustness:
While baselines exhibit volatility (SAND degrades 5.8 times on WSD vs. NAB), \name consistently excels in F1-E, ranking 1st in all datasets. This highlights \name’s efficacy for complex real-world scenarios and validates strong generalization beyond dataset-specific biases.

Advantage in Noisy Environments:
On AIOPS, where infrastructure drift and heterogeneous anomalies challenge detectors, \name F1-E (0.8049) surpasses LSTMAD (0.7671) and FCVAE (0.7364). This suggests superior handling of non-stationary patterns and rare anomaly types.

Our method sets a new state-of-the-art in time series anomaly detection, particularly for dynamic systems with complex local shapes. The consistent gains across metrics and datasets affirm its suitability for operational deployments.

\subsection{Ablation Study}
We validate the effectiveness of each technique in \name through multiple ablation studies as follows.

\subsubsection{Effectiveness of Multi-scale Convolution}
\begin{figure}[tbp]
  \centering
  \includegraphics[width=0.47\textwidth]{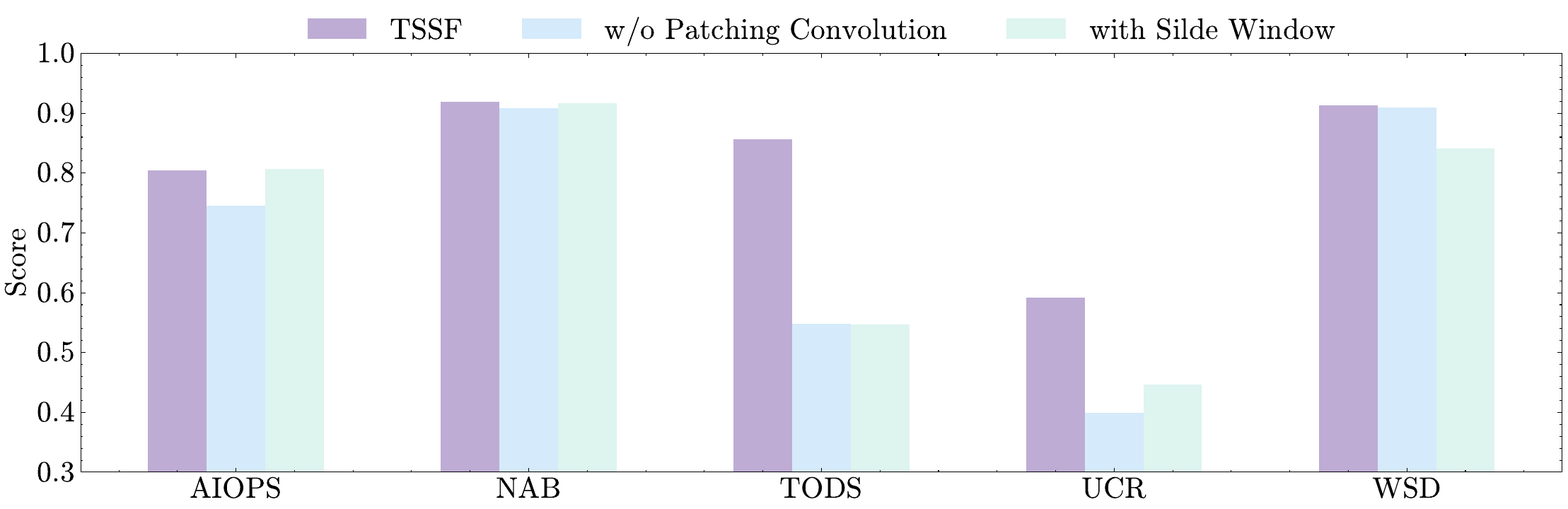}
  \caption{ Effectiveness Evaluation of Multi-scale Convolution }
  \label{figure:Ablation_patch}
  \vspace{-0.3cm}
\end{figure}

To quantify the contribution of our Multi-scale Convolution module, we conduct two ablation variants: \textbf{w/o Patch-wise Convolution}: remove all convolution feature extractors and directly pass raw patches to the attention layers; \textbf{with Sliding Window}: replace multi‑scale convolution with sliding‑window average pooling over each patch. Fig.\ref{figure:Ablation_patch} reports the Event‑F1 scores for our full model and the two variants across the five benchmarks.

Removing the Multi-scale Convolution block incurs a substantial performance drop on TODS and UCR.  This indicates that multi‑scale convolution is critical for extracting discriminative local patterns, especially in complex settings.  The Sliding Window variant fails to match our convolutional design on TODS and UCR, confirming that learned filters provide richer, scale‑adaptive descriptors than fixed pooling.  These results validate that Multi-scale Convolution is a necessary component for robust anomaly detection across diverse time‑series environments.

\subsubsection{Effectiveness of Patch-wise Dual-Attention mechanism}

\begin{figure}[tbp]
  \centering
  \includegraphics[width=0.45\textwidth]{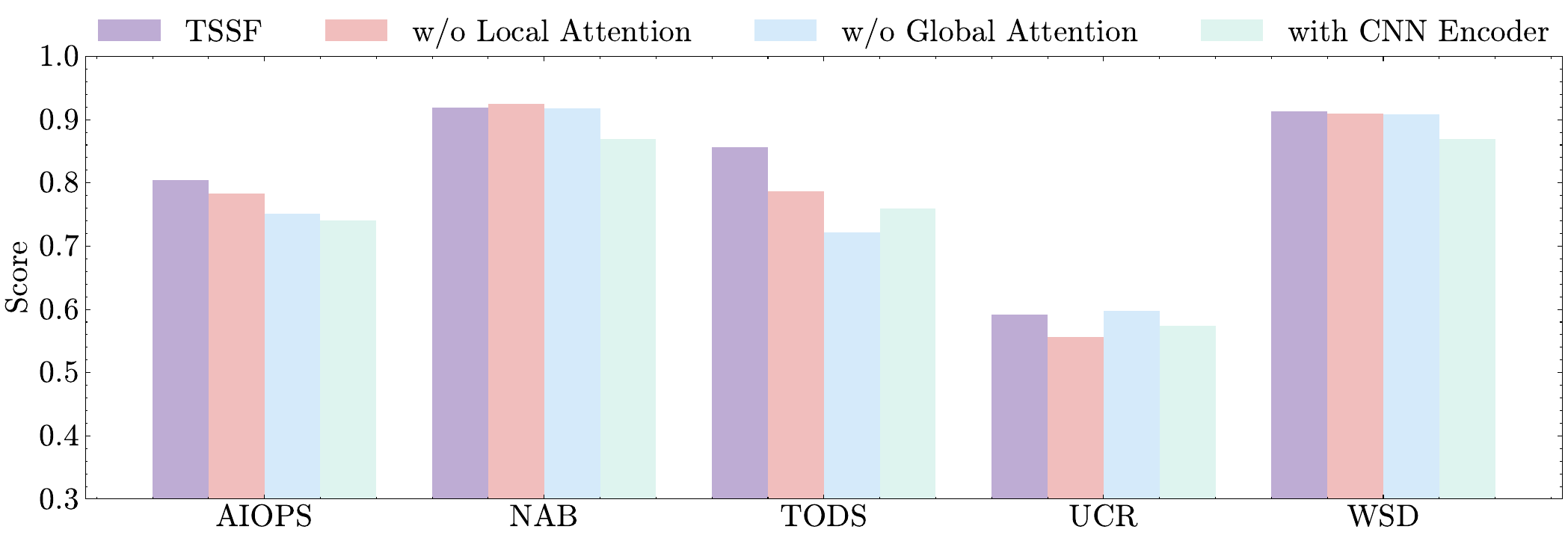}
  \caption{ Effectiveness Evaluation of Patch-wise Dual‑Attention }
  \label{figure:ablation_attention}
  \vspace{-0.3cm}

\end{figure}

To evaluate the impact of our dual-attention design, we compare against three ablation variants: \textbf{w/o Local Attention}: Global attention over patches only. \textbf{w/o Global Attention}: Local attention within each patch only. \textbf{CNN Encoder}: Replaces dual-attention with a TimesNet-style~\cite{wu2022timesnet} convolution encoder. Fig.\ref{figure:ablation_attention} presents the performance metrics for \name\ and each variant. On the NAB and WSD datasets, where anomalies exhibit relatively simplistic patterns, both local-only and global-only attention configurations demonstrate competent detection performance, resulting in unstable differentiation between variants. However, these ablation studies collectively confirm that both attention streams are essential for robust anomaly detection across diverse time series.

\begin{figure}[htbp]
  \centering
  \includegraphics[width=0.45\textwidth]{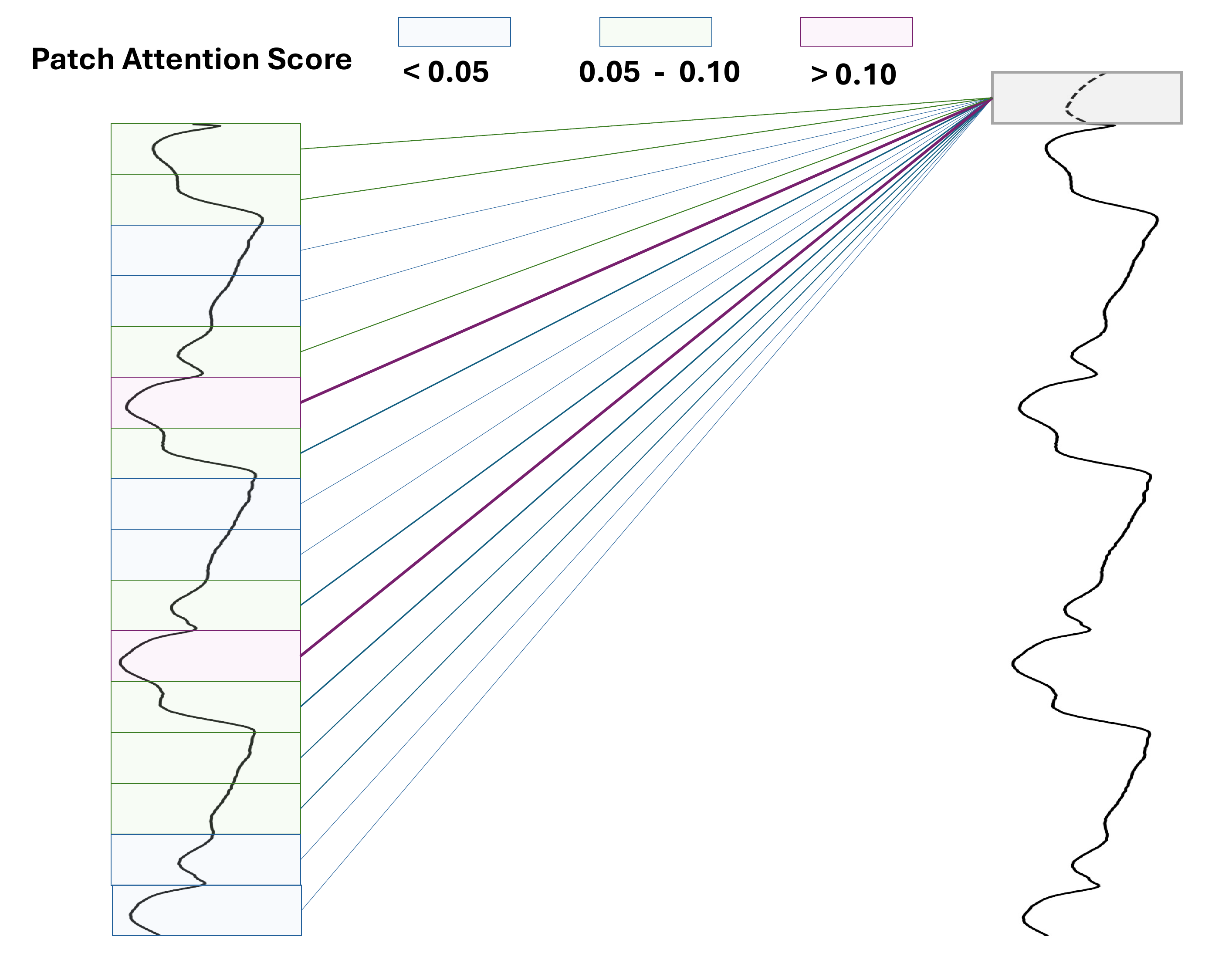}
  \caption{ Visualization of patch attention score }
  \label{figure:attention score}
\end{figure}

\subsection{Visualization of patch attention score}
To illustrate how \name leverages both local and global contexts when an anomaly occurs, we visualize the patch‑level attention scores over a representative anomalous interval UCR InternalBleeding10. Fig.\ref{figure:attention score} plots the normalized attention weights assigned to each of the patches at the time step. 
\textbf{Global Context:} Non‑adjacent patches with similar temporal patterns receive the highest attention scores, demonstrating that \name captures long‑range dependencies and recurring motifs.
\textbf{Localized Focus:} Patches whose time windows overlap the anomaly also exhibit elevated attention, indicating that the model correctly emphasizes the segments containing true anomalies.  
\textbf{Smooth Distribution:} The attention profile decays smoothly across distant patches, avoiding sharp drop that could lead to missing related contextual cues. This visualization confirms that \name dual‑attention mechanism produces a coherent, correct distribution of importance across patches, validating its ability to integrate fine‑grained and long‑range information when detecting anomalies.

%% file: Conference-LaTeX-template_10-17-19/chapters/01-content/50-CONCLUSION.tex
\section{CONCLUSION}\label{sec:CONCLUSION}

We have introduced \name, a novel framework that combines Multi-scale Convolution with a dual‐attention mechanism and learnable gating to capture both fine‐grained local shapes and long‐range dependencies in univariate time series.  Extensive experiments on five diverse benchmarks show that \name achieves a new state‐of‐the‐art average score, outperforming leading methods by over 10\% and demonstrating strong robustness in noisy, non‑stationary environments.  Ablation studies confirm the critical role of each component, while visualization of patch attention highlights its coherent integration of local and global contexts. 

%% file: Conference-LaTeX-template_10-17-19/chapters/01-content/Acknowledgement.tex
\section{Acknowledgement}\label{sec:Acknowledgement}

This work was funded by the National Natural Science Foundation
of China (62202445), and the National Natural Science Foundation
of China-Research Grants Council (RGC) Joint Research Scheme
(62321166652).

%% file: Conference-LaTeX-template_10-17-19/issre2025.bbl
\begin{thebibliography}{10}
\providecommand{\url}[1]{#1}
\csname url@samestyle\endcsname
\providecommand{\newblock}{\relax}
\providecommand{\bibinfo}[2]{#2}
\providecommand{\BIBentrySTDinterwordspacing}{\spaceskip=0pt\relax}
\providecommand{\BIBentryALTinterwordstretchfactor}{4}
\providecommand{\BIBentryALTinterwordspacing}{\spaceskip=\fontdimen2\font plus
\BIBentryALTinterwordstretchfactor\fontdimen3\font minus \fontdimen4\font\relax}
\providecommand{\BIBforeignlanguage}[2]{{%
\expandafter\ifx\csname l@#1\endcsname\relax
\typeout{** WARNING: IEEEtran.bst: No hyphenation pattern has been}%
\typeout{** loaded for the language `#1'. Using the pattern for}%
\typeout{** the default language instead.}%
\else
\language=\csname l@#1\endcsname
\fi
#2}}
\providecommand{\BIBdecl}{\relax}
\BIBdecl

\bibitem{liu2025opseval}
Y.~Liu, C.~Pei, L.~Xu, B.~Chen, M.~Sun, Z.~Zhang, Y.~Sun, S.~Zhang, H.~Zhang, G.~Xie \emph{et~al.}, ``Opseval: A comprehensive benchmark suite for evaluating large language models’ capability in it operations domain,'' 2025.

\bibitem{Microsoft}
Microsoft, ``Microsoft's azure networking takes a worldwide tumble,'' \url{https://www.theregister.com/2024/07/30/microsofts_azure_portal_outage/}, 2025.

\bibitem{Google}
Google, ``Google cloud services hit by outage in paris,'' \url{https://thenewstack.io/google-cloud-services-hit-by-outage-in-paris/}, 2023.

\bibitem{Alibaba}
Alibaba, ``Alibaba cloud health dashboard,'' \url{https://status.aliyun.com/#/historyEvent}, 2025.

\bibitem{si2024timeseriesbench}
H.~Si, J.~Li, C.~Pei, H.~Cui, J.~Yang, Y.~Sun, S.~Zhang, J.~Li, H.~Zhang, J.~Han \emph{et~al.}, ``Timeseriesbench: An industrial-grade benchmark for time series anomaly detection models,'' in \emph{2024 IEEE 35th International Symposium on Software Reliability Engineering (ISSRE)}.\hskip 1em plus 0.5em minus 0.4em\relax IEEE, 2024, pp. 61--72.

\bibitem{xu2021AnomalyTransformer}
J.~Xu, H.~Wu, J.~Wang, and M.~Long, ``Anomaly transformer: Time series anomaly detection with association discrepancy,'' \emph{arXiv preprint arXiv:2110.02642}, 2021.

\bibitem{wang2024revisiting}
Z.~Wang, C.~Pei, M.~Ma, X.~Wang, Z.~Li, D.~Pei, S.~Rajmohan, D.~Zhang, Q.~Lin, H.~Zhang \emph{et~al.}, ``Revisiting vae for unsupervised time series anomaly detection: A frequency perspective,'' in \emph{Proceedings of the ACM Web Conference 2024}, 2024, pp. 3096--3105.

\bibitem{yan2022shapeformer}
X.~Yan, L.~Lin, N.~J. Mitra, D.~Lischinski, D.~Cohen-Or, and H.~Huang, ``Shapeformer: Transformer-based shape completion via sparse representation,'' in \emph{Proceedings of the IEEE/CVF conference on computer vision and pattern recognition}, 2022, pp. 6239--6249.

\bibitem{lu2022DAMP}
Y.~Lu, R.~Wu, A.~Mueen, M.~A. Zuluaga, and E.~Keogh, ``Matrix profile xxiv: scaling time series anomaly detection to trillions of datapoints and ultra-fast arriving data streams,'' in \emph{Proceedings of the 28th ACM SIGKDD conference on knowledge discovery and data mining}, 2022, pp. 1173--1182.

\bibitem{zhong2025patchad}
Z.~Zhong, Z.~Yu, Y.~Yang, W.~Wang, K.~Yang, and C.~P. Chen, ``Patchad: A lightweight patch-based mlp-mixer for time series anomaly detection,'' \emph{IEEE Transactions on Big Data}, 2025.

\bibitem{breunig2000lof}
M.~M. Breunig, H.-P. Kriegel, R.~T. Ng, and J.~Sander, ``Lof: identifying density-based local outliers,'' in \emph{Proceedings of the 2000 ACM SIGMOD international conference on Management of data}, 2000, pp. 93--104.

\bibitem{boniol2021sand}
P.~Boniol, J.~Paparrizos, T.~Palpanas, and M.~J. Franklin, ``Sand: streaming subsequence anomaly detection,'' \emph{Proceedings of the VLDB Endowment}, vol.~14, no.~10, pp. 1717--1729, 2021.

\bibitem{zhu2018matrix}
Y.~Zhu, C.-C.~M. Yeh, Z.~Zimmerman, K.~Kamgar, and E.~Keogh, ``Matrix profile xi: Scrimp++: time series motif discovery at interactive speeds,'' in \emph{2018 IEEE international conference on data mining (ICDM)}.\hskip 1em plus 0.5em minus 0.4em\relax IEEE, 2018, pp. 837--846.

\bibitem{rousseeuw2003AR}
P.~J. Rousseeuw and A.~M. Leroy, \emph{Robust regression and outlier detection}.\hskip 1em plus 0.5em minus 0.4em\relax John wiley \& sons, 2003.

\bibitem{malhotra2015LSTMAD}
P.~Malhotra, L.~Vig, G.~Shroff, P.~Agarwal \emph{et~al.}, ``Long short term memory networks for anomaly detection in time series,'' in \emph{Proceedings}, vol.~89, no.~9, 2015, p.~94.

\bibitem{wu2022timesnet}
H.~Wu, T.~Hu, Y.~Liu, H.~Zhou, J.~Wang, and M.~Long, ``Timesnet: Temporal 2d-variation modeling for general time series analysis,'' \emph{arXiv preprint arXiv:2210.02186}, 2022.

\bibitem{zhou2023OFA}
T.~Zhou, P.~Niu, L.~Sun, R.~Jin \emph{et~al.}, ``One fits all: Power general time series analysis by pretrained lm,'' \emph{Advances in neural information processing systems}, vol.~36, pp. 43\,322--43\,355, 2023.

\bibitem{ng2011AE}
A.~Ng \emph{et~al.}, ``Sparse autoencoder,'' \emph{CS294A Lecture notes}, vol.~72, no. 2011, pp. 1--19, 2011.

\bibitem{malhotra2016EncDecAD}
P.~Malhotra, A.~Ramakrishnan, G.~Anand, L.~Vig, P.~Agarwal, and G.~Shroff, ``Lstm-based encoder-decoder for multi-sensor anomaly detection,'' \emph{arXiv preprint arXiv:1607.00148}, 2016.

\bibitem{tuli2022tranad}
S.~Tuli, G.~Casale, and N.~R. Jennings, ``Tranad: Deep transformer networks for anomaly detection in multivariate time series data,'' \emph{arXiv preprint arXiv:2201.07284}, 2022.

\bibitem{AIOPS}
``Aiops competition,'' 2018, \url{https://github.com/huggingface/candle}.

\bibitem{WSD}
S.~Zhang, Z.~Zhong, D.~Li, Q.~Fan, Y.~Sun, M.~Zhu, Y.~Zhang, D.~Pei, J.~Sun, Y.~Liu \emph{et~al.}, ``Efficient kpi anomaly detection through transfer learning for large-scale web services,'' \emph{IEEE Journal on Selected Areas in Communications}, vol.~40, no.~8, pp. 2440--2455, 2022.

\bibitem{UCR}
R.~Wu and E.~J. Keogh, ``Current time series anomaly detection benchmarks are flawed and are creating the illusion of progress,'' \emph{IEEE transactions on knowledge and data engineering}, vol.~35, no.~3, pp. 2421--2429, 2021.

\bibitem{TODS}
K.-H. Lai, D.~Zha, J.~Xu, Y.~Zhao, G.~Wang, and X.~Hu, ``Revisiting time series outlier detection: Definitions and benchmarks,'' in \emph{Thirty-fifth conference on neural information processing systems datasets and benchmarks track (round 1)}, 2021.

\bibitem{NAB}
S.~Ahmad, A.~Lavin, S.~Purdy, and Z.~Agha, ``Unsupervised real-time anomaly detection for streaming data,'' \emph{Neurocomputing}, vol. 262, pp. 134--147, 2017.

\bibitem{ren2019SRCNN}
H.~Ren, B.~Xu, Y.~Wang, C.~Yi, C.~Huang, X.~Kou, T.~Xing, M.~Yang, J.~Tong, and Q.~Zhang, ``Time-series anomaly detection service at microsoft,'' in \emph{Proceedings of the 25th ACM SIGKDD international conference on knowledge discovery \& data mining}, 2019, pp. 3009--3017.

\bibitem{zhang2022tfad}
C.~Zhang, T.~Zhou, Q.~Wen, and L.~Sun, ``Tfad: A decomposition time series anomaly detection architecture with time-frequency analysis,'' in \emph{Proceedings of the 31st ACM international conference on information \& knowledge management}, 2022, pp. 2497--2507.

\bibitem{Metric_Best_f1}
H.~Xu, W.~Chen, N.~Zhao, Z.~Li, J.~Bu, Z.~Li, Y.~Liu, Y.~Zhao, D.~Pei, Y.~Feng \emph{et~al.}, ``Unsupervised anomaly detection via variational auto-encoder for seasonal kpis in web applications,'' in \emph{Proceedings of the 2018 world wide web conference}, 2018, pp. 187--196.

\bibitem{xu2023fits}
Z.~Xu, A.~Zeng, and Q.~Xu, ``Fits: Modeling time series with $10 k $ parameters,'' \emph{arXiv preprint arXiv:2307.03756}, 2023.

\bibitem{Best_f1_inflation}
R.~Wu and E.~J. Keogh, ``Current time series anomaly detection benchmarks are flawed and are creating the illusion of progress,'' \emph{IEEE transactions on knowledge and data engineering}, vol.~35, no.~3, pp. 2421--2429, 2021.

\end{thebibliography}
